\documentclass[preprint2]{aastex}
\input{psfig.sty}
\input{loop1.sty}
\begin{document}
\title{\bf Pulsar Scintillation in the Local ISM: Loop I and Beyond}
\author{N. D. Ramesh Bhat}
\affil{National Astronomy and Ionosphere Center, Arecibo Observatory, 
HC 3 Box 53995, PR 00612, USA.}
\author{Yashwant Gupta}
\affil{National Centre for Radio Astrophysics, Tata Institute of 
Fundamental Research, Pune 411007, India.}
\begin{abstract}
Recent pulsar scintillation measurements from Ooty, in conjunction with 
those from Parkes and other large radio telescopes, are used for a 
systematic investigation of the Local Interstellar Medium (LISM) towards 
the general direction of the Loop I Bubble.
For several pulsars, clear evidence is found for an enhanced level of 
scattering which is over and above what can be accounted for by the 
enhanced scattering model for the Local Bubble. 
These results are interpreted in terms of enhanced scattering due to 
turbulent plasma associated with the Loop I shell.
Useful constraints are obtained for the scattering properties of the shell.
The inferred value for the scattering measure for the Loop I shell is 
found to be $\sim$0.3 \smu.
Assuming a shell thickness $\sim$5--10 pc, this implies an average strength 
of scattering in the shell that is $\sim$100--200 times larger 
than that in the ambient ISM. An alternative explanation, where the enhanced
level of scattering is due to a possible ``interaction zone'' between the 
Local Bubble and the Loop I Bubble, is also considered; it is found to be 
somewhat less satisfactory in explaining the observations.  The best fit 
value for the scattering measure for such an interaction zone region is 
estimated to be $\sim$1.1 \smu.

Further, several pulsars beyond $\sim$1 kpc are found to show enhanced levels 
of scattering over and above that expected from this ``two-bubble model.'' 
For some of the low-latitude pulsars, this is found to be due to enhanced 
scattering from plasma inside the intervening Sagittarius spiral arm.
We discuss the implications of our results for the interpretation of 
scintillation data and for the general understanding of the LISM.
\end{abstract}


\bigskip

\section{Introduction}
\label{s:intro}

The Interstellar Medium (ISM) within a few hundred parsecs of the Solar System 
has been a topic of considerable observational and theoretical investigation 
over recent years (e.g. Breitschwerdt, Freyberg \& Tr\"umper 1998; Frisch 1996; 
Cox \& Reynolds 1987). 
This region, often referred to as the Local Interstellar Medium (LISM), is 
known to contain several major features in the form of bubbles, supernova 
shells and \HI clouds.
The Solar System itself is thought to reside in a low-density, X-ray--emitting 
cavity of a mean radius $\sim$ 100 pc, a region usually referred to as the Local Bubble
(e.g. Cox \& Reynolds 1987). It is believed to be the remnant of a supernova 
explosion that occurred $\sim 10^7$ yr ago.
Prominent amongst other features in the LISM are the four large-diameter, almost circular 
rings seen in the all-sky distribution of radio continuum emission.
These are appropriately named Radio Loops I to IV (Haslam, Khan \& Meaburn 1971)
and are thought to be the projections of quasi-spherical bubbles.
Of these, the Loop I Bubble---the largest in angular extent (solid angle 
$\sim {7 \over 6} \pi$ steradian) and also the brightest of the 
four loops---is of special interest (see Salter (1983) for a review).
Its close proximity to the Sun allows an in-depth study; see for example, 
the work of Egger (1993) using data from the ROSAT soft X-ray Sky Survey, and 
that of Nishikida (1999) who combined the ROSAT PSPC data with IRAS Sky 
Survey and radio 21-cm data. 
In terms of origin, Loop I is thought to be an expanding super-bubble triggered 
by an epoch of star formation in the Sco-Cen association some $\sim10^6$ years ago.
Further, it has been postulated that the properties of the LISM 
may well be conditioned by the outer shock wave of this supernova remnant
(Frisch 1981, 1996; Lallement 1998). 
Besides these radio loops, the other well-known examples of nearby bubbles 
that may be important in understanding the LISM are: the Eridanus 
bubble (distance from the Sun, D $\sim$ 100--150 pc), the Gum Nebula 
(D$\sim$200--250 pc), the Orion Bubble (D $\sim$ 455 pc) and the 
Monogem ring (D $\sim$ 100--1300 pc). 

Recent X-ray and UV data from ROSAT have led to several new 
insights into the structure of the LISM. A noteworthy result comes from ROSAT 
PSPC data suggesting an ongoing interaction between the Local Bubble and the
Loop I Bubble. 
By comparing the diffuse X-ray background maps (in the 0.1--2.0 keV band) from 
the ROSAT All-Sky Survey (Tr\"umper 1983) and \HI data over a 
$160^{\circ} \times 160^{\circ}$ region centered at 
($l,b$)=($329^{\circ},+17.5^{\circ}$; the apparent centre of the Radio 
Loop I), Egger \& Aschenbach (1995; hereafter EA95; see also Egger 1993, 
1998) recognize a ``ring-like'' structure between the Local Bubble and 
Loop I, where the \HI column density (\NH) is even higher than that of the 
intervening dense \HI shell known to exist in the direction of the Sco-Cen OB 
association (Centurion \& Vladilo 1991). 
From absorption-line studies of nearby stars, the distance to this 
interaction feature (where \NH jumps by nearly an order of magnitude)
is estimated to be $\sim$ 70 pc (see EA95),
quite comparable to the distance to the neutral \HI wall ($\sim 40 \pm 25$ pc) 
inferred from ROSAT WFC star counts (Warwick et al. 1993) and optical and UV 
spectral line data (Centurion \& Vladilo 1991). 
Interestingly, the formation of such a ring and the inferred density 
enhancement (by a factor $\sim$ 25) are in good agreement with the numerical 
simulations of colliding interstellar bubbles (Yoshioka \& Ikeuchi 1990).
In this picture, Loop I is considered to be an active super-bubble, 
with at least one bubble having already formed a dense cool shell prior to 
collision. 
However, alternative interpretations do exist (e.g., Frisch 1996, 1998),
wherein the Local Bubble is pictured as an appendix of Loop I.
More data and modeling may help to distinguish clearly between 
the different scenarios.

While much of our understanding of the structure of the LISM comes primarily
from X-ray and UV data, nearby pulsars are promising tools for enhancing 
this understanding.
Studies of dispersion and scattering of the radio signals from pulsars are 
useful means for probing the intervening ISM.  Of these two, interstellar 
scintillation (ISS) effects are more likely to be influenced by the peculiar 
distribution of the ionized plasma along the line of sight (LOS) --- such as 
clumps of enhanced density superposed on a uniform distribution of ionized 
material.
This is mainly because of (a) the nonlinear relation(s) of the scintillation
properties to the electron density, and (b) the fact that scintillation 
effects depend critically on the relative location of the scatterer (or more generally,
the actual distribution of scattering plasma along the sight line).
For instance, if the medium is inhomogeneous, 
in order to produce noticeable effects in the dispersion measure (DM), 
the density at the clumps has to be considerably larger 
than to produce similar effects in the scintillation data.
However, if the pulsar happens to lie within or near the clumped region,
its effect on the scintillation properties could be significantly reduced.
Thus, in some sense, dispersion and scintillation data can provide 
information complementary to each other.
Nevertheless, (b) would imply that the interpretation of scintillation data 
is less immune to distance errors, and hence offers a better handle.

It is quite plausible that the distribution of ionized plasma in and around
large local features such as the Local Bubble and Loop I can considerably 
influence the dispersion and scintillation of nearby pulsars, and in some 
cases even scintillation of extra-galactic radio sources.  
The large-scale distribution of free electrons and turbulent
scattering plasma in the Galaxy has been modeled
by Taylor \& Cordes (1993; hereafter TC93).  While sophisticated compared 
to its predecessors, the TC93 model takes very little account of the peculiar 
properties of the LISM due to individual interstellar features. 
The recent years have seen an accumulation of observational evidence for
the effects of such features in the LISM.
Early investigations include the work of Phillips \& Clegg (1992), who
proposed that the scattering of radiation from the nearby pulsar PSR 
B0950+08 is probably dominated by weakly turbulent plasma 
present in the interior of the Local Bubble, and that of Hajivassiliou 
(1992), who invoked an ellipsoidal envelope of highly turbulent plasma to explain 
the directional anisotropy seen in the turbulent intensity maps derived 
from interplanetary scintillation studies of radio sources. 
Even earlier, Rickard \& Cronyn (1979) had suggested scattering 
from the outer shell of Loop I as the plausible cause of 
a statistically significant lack of interplanetary scintillators seen 
in a band some $20^{\circ}$ outside the brightest section of the 
Loop I radio emission, the North Polar Spur (NPS).

In a previous paper (Bhat, Gupta \& Rao 1998; hereafter BGR98), we presented a 
detailed, systematic study of the LISM using
pulsar scintillation data from the Ooty Radio Telescope (ORT). Our results and
analysis strongly support the view that the scattering in the LISM is probably 
dominated by turbulent plasma at the boundaries of the Local Bubble.
We proposed a simple model, wherein the Solar system is surrounded 
by an ellipsoidal shell-like structure, with a size of $\sim$100 pc in the 
Galactic plane and $\sim$500 pc in a plane perpendicular to this
(Fig.~\ref{plotfour}).
The scattering structure has its center located at $\sim$20--35 pc 
from the Sun towards 
$215^{\circ} < l < 240^{\circ}$, $-20^{\circ} < b < +20^{\circ}$.
In this picture, the interior of the bubble is filled with plasma of 
relatively low turbulence (characterized by a scattering strength,
~\avcn~ $\sim 10^{-4}$ \cnu), whereas the shell material
(thickness $\sim$1--10 pc) has a scattering strength $\sim$50--800 times 
larger than that in the ambient ISM (integrated strength of scattering,
scattering measure, given by 0.11$<$\SMLB$<$0.28 \smu).
The contribution of the shell thus dominates the total 
scattering, which would imply that the scattering geometry towards many 
sight lines can be approximated by a ``thin screen'' placed at the bubble boundary
(location in the range $\sim$20--200 pc).
This model successfully explained the enhanced level of scattering measured
towards a number of nearby pulsars.

The Ooty experiment covered only a few pulsars that would be useful
for a study of Loop I.
However, the recent results from Parkes observations of a large number of southern 
pulsars (Johnston, Nicastro \& Koribalski 1998) has significantly improved 
the ISS data available for probing the ISM in and around Loop I.  This 
motivated us to take a more detailed look at the distribution of scattering 
material towards Loop I and beyond, a study which forms the main theme of this paper.
The paper is organized as follows: 
\S~\ref{s:data} describes observational data and our analysis techniques; 
\S~\ref{s:data-loop1} and \S~\ref{s:data-spiral} present the modeling of 
the scattering plasma associated with Loop I. 
The uncertainties relevant to our analysis are discussed in 
\S~\ref{s:res-loop1}. 
In \S~\ref{s:res-zone}, we consider some possible alternative models, while
later sections of \S~\ref{s:res} discuss some general implications of our 
results for scintillation data and for the LISM.
In \S~\ref{s:conc}, we summarize our conclusions.

\section{Observational Data, Analysis Techniques and Modeling}
\label{s:data}

\subsection{Sample Selection}
\label{s:data-sample}

For the present analysis, we are interested in pulsars whose scintillation 
properties are likely to be influenced by the structure of Loop I.
In order to pre-select pulsars useful for this purpose, we 
adopt a geometry and size for Loop I from the published 
literature.
Specifically, Berkhuijsen, Haslam \& Salter (1971) derived an angular 
diameter of $116^{\circ} \pm 4^{\circ}$, and a centre at 
$(l,b)=(329.0^{\circ} \pm 1.5^{\circ},+17.5^{\circ} \pm 3.0^{\circ}$; 
see also EA95).
Further, on the basis of the plausible connection between the origin
of Loop I and the Sco-Cen OB association (e.g. Egger
\& Aschenbach 1995), it is fair to assume the loop center 
to be near the center of mass of the association ($\sim$170 pc).
Based on the above, we model the Loop I Bubble as a spherical 
shell of size $\sim$290 pc, with the center located at 
$\sim$170 pc towards ($329^{\circ}, 17.5^{\circ}$).
This yields 52 pulsars within $\sim$ 2 kpc of the Sun whose sight lines 
intersect the projected area of Loop I on the sky. 
On carrying out a literature search, we find scintillation measurements
to be available for only 20 of these (see Table~\ref{tableone}).
The majority of these scintillation measurements are from recent observations 
with the Parkes and Ooty radio telescopes, while a few are from Arecibo 
(data from Johnston et al. 1998; Bhat et al. 1999b, Gothoskar \& Gupta 
2000, and Cordes 1986).
The measurements of decorrelation bandwidth (\ndmeas) and the observing 
frequencies (\fobs) are listed in columns (5) and (6) of Table~\ref{tableone},
respectively.
The distance estimates in column (4) are derived from dispersion measures (DMs) and the 
model of Galactic electron density (\nele) by TC93, except for PSRs J1456--6843 
and J1744--1134.  For these two pulsars, we use independent distance estimates 
derived from the measurement of annual trigonometric parallax (Bailes et al. 
1990; Toscano et al. 1999).

Although the decorrelation bandwidth measurements listed in 
Table~\ref{tableone} are at different observing frequencies, we 
have scaled them to a common frequency of 327 MHz, assuming a 
Kolmogorov scaling law (i.e., a wavenumber spectrum with a slope 
of 11/3 over the spatial scales of interest; 
$\nd \propto {\rm frequency}^{4.4}$).
We note that this can potentially produce some errors, as the exact nature of 
the electron density wavenumber spectrum is still debated.
However, there is substantial observational 
evidence in favor of an $\alpha \approx 11/3$ spectrum towards many sight
lines in the LISM (e.g. Bhat, Gupta \& Rao 1999a).
Even if this were not strictly correct, most measurements from Parkes 
will be only marginally affected by an incorrect frequency scaling. 
This scaling bias may be significant for \nd values at 1.5 GHz; however,
these are for relatively distant objects (D$>$1.2 kpc), and hence less 
critical for the investigation of scattering in the LISM.

\subsection{Distribution of Scattering: Choice of the Method}
\label{s:data-method}

There are three different methods by which 
pulsar scintillation measurements can be used to investigate 
the distribution of scattering material in the Galaxy:
(1) using the decorrelation bandwidth, \nd (or its equivalent, 
the temporal pulse broadening time, \taup) --- this quantifies 
the scattering measure (SM) which characterizes the total amount of 
scattering along the line of sight. 
This technique has been used extensively to investigate the large-scale
distribution of \cn in the Galaxy (Cordes, Weisberg \& Boriakoff 1985;
CWB85 hereafter; Cordes et al. 1991; TC93):
(2) using measurements of angular broadening (\tscatt) in conjunction with 
\taup --- wherein the differing weighting functions of the two observables 
can be used to determine a more exact distribution of \cn along the LOS
(e.g. Gwinn, Bartel \& Cordes 1993):
(3) using the hybrid method recently proposed by 
Cordes \& Rickett (1998) --- wherein the diffractive scintillation measurements 
(decorrelation bandwidth and scintillation timescale, \td), in conjunction 
with the pulsar proper motion and distance, can be used to obtain the 
distribution of \cn along the LOS (e.g. Chatterjee et al. 2000). 

Of these methods, (2) is not relevant in our case, as measurements of 
\tscatt and \taup do not exist for bulk of the objects in Table~\ref{tableone}.
To date, proper motion measurements are known for 9 of the 20 pulsars in 
Table~\ref{tableone}.
However, uncertainties are too large for the values to be useful in most 
cases, and we were unable to derive any meaningful results on the
distribution of scattering in and around Loop I.
We therefore restricted ourselves to method (1) for the analysis described 
in this paper (see Appendix A for a detailed description of the method).

\subsection{Comparison with the Local Bubble model}
\label{s:data-comp}

\begin{figure}
\hskip 0.5in
\psfig{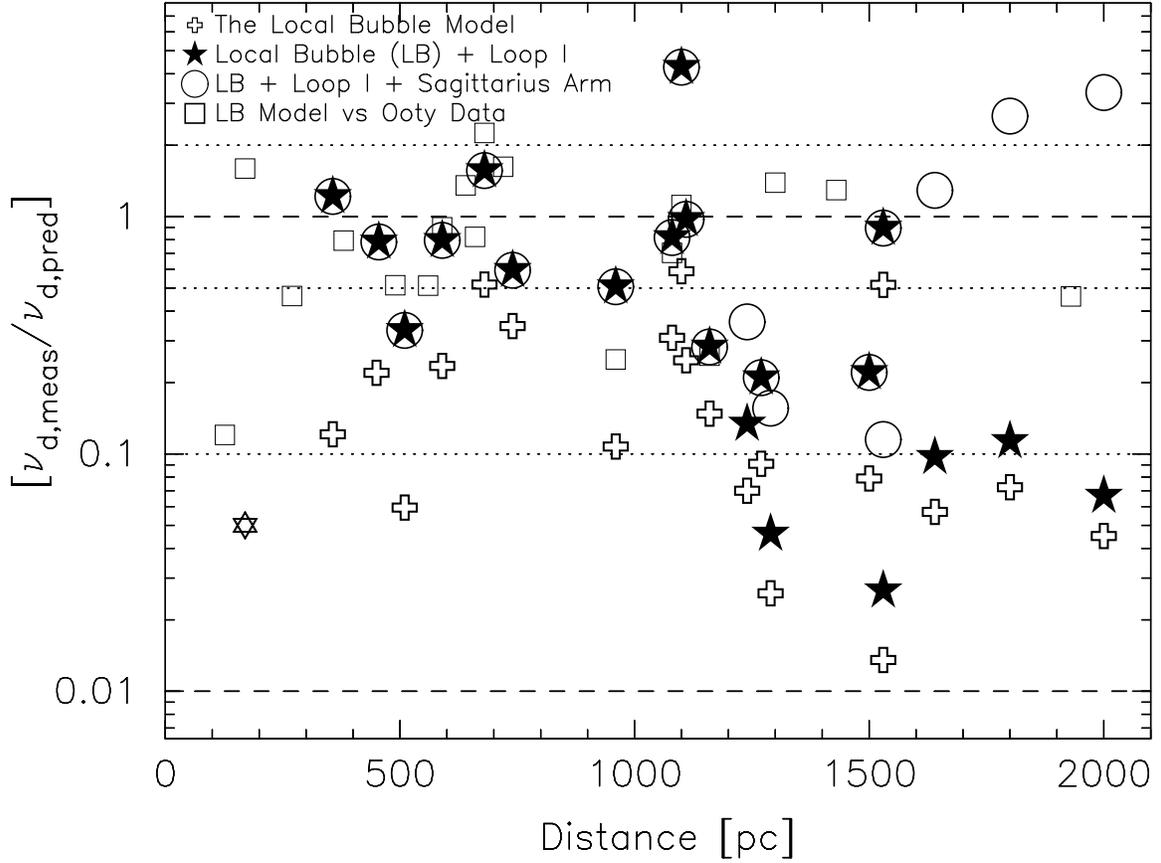}
\caption[]{Ratios of the measured decorrelation bandwidths (\ndmeas) to their 
predictions from the various models for the distribution of scattering in the 
Local ISM (\ndpred) are plotted against the distance estimates. The \nd values 
are scaled to a common frequency of 327 MHz. The lone unfilled star indicates 
the measurement of PSR J1744--1134 at its TC93 distance of 166 pc.}
\label{plotone}
\end{figure}

We first examine how well the scintillation data in Table~\ref{tableone} agree 
with the predictions of the Local Bubble model of BGR98, as described
in \S~\ref{s:intro}.
Fig.~\ref{plotone} shows a plot of the ratio of the measured to predicted 
decorrelation bandwidths (\ndratio) against the pulsar distance estimates.
We introduce a quantity, $\epsA$, that is a measure of the degree of agreement 
between \ndmeas and \ndpred. This is expressed as 

\begin{equation}
\epsA = { 1 \over N_p - 1 } \sum _{i=1} ^{i=N_p} \left [ {\rm log} 
\left ( { \ndmeas \over \ndpred} \right ) _i ~\right ]^2 
\end{equation}

\noindent
where $N_p$ is the total number of pulsars for which the comparison is being
made. The logarithm has been
taken to give equal weight to discrepancies that are below and above 
unity when computing 
\epsA . \footnote{We will use the quantities \epslow and \epshigh when
referring to those parts of the data that correspond to objects with D \la 1 
kpc and D \ga 1 kpc, respectively.}

As is obvious from Fig.~\ref{plotone}, there are large discrepancies (ranging from 
a factor 2 to as much as $\sim$50) for the case of the Local Bubble model
(shown by crosses).  
Considering that most measurements in Table~\ref{tableone} are {\it not} from 
observations averaged over many epochs, a potential explanation for part of
these discrepancies may be errors due to long-term refractive interstellar 
scintillation (RISS) effects (e.g. Gupta, Rickett \& Lyne 1994; 
Bhat et al. 1999a). 
However, most discrepancies are considerably larger than the worst 
case factor of 3--5 that can be accounted for by RISS effects.
Moreover, it is striking that all the ratios are less than unity, implying that 
the scattering strengths are systematically larger than that can be accounted 
for by the Local Bubble model. 
Further, a systematic trend with distance is also evident in Fig.~\ref{plotone}, 
whereby \ndratio~ (for the Local Bubble model) tends to lie in the range 0.01--0.1 for pulsars beyond 1.2 kpc, 
but between 0.1 and 1 for those nearer than this. 
In contrast, for a sample of pulsars within 1 kpc of the Sun in the 
complementary sky (shown by $\Box$ in Fig.~\ref{plotone}, cf. BGR98), 
most ratios fall within a factor 2--3 of unity.
The value of $\epsA$ for this sample ($\approx$~0.07) is about 16 times lower 
than that for the Local Bubble model applied to the current data set 
($\epsA \approx 1.1$, as shown in Table~\ref{tablethree}).
The simplest interpretation is that there is yet another source of excess 
scattering (in addition to the Local Bubble shell) in the general 
direction of Loop I.

\subsection{Enhanced Scattering from the Shell of the Loop I Bubble?}
\label{s:data-loop1}

A number of previous studies have revealed the existence of strong excess 
scattering towards several sight lines in the Galaxy (e.g. Moran et al.
1990).
In the context of modeling the large-scale distribution of \cn~in the Galaxy, 
Cordes et al. (1991) 
propose a `clump' component, possibly associated with supernova shocks or \HII 
regions, to explain such unusually high scattering. 
These are regions of intense turbulence, with the strength of scattering 
many orders of magnitude larger than that in the typical ISM.
However, the mean free paths expected for such clumps are 
rather large (typically $\sim$5--10 kpc).
There seems to be no
observational evidence for the presence of any such regions in the LISM.
Strong excess scattering may also arise if the sight line to a pulsar 
intercepts an \HII region (e.g. Gupta et al. 1994). 
To the best of our knowledge, none of the pulsars in Table~\ref{tableone} 
represents such a situation.
Even if the LISM contains strong scattering regions that are hitherto
unknown, these could be relevant only for a few lines-of-sight,
whereas the data in Fig.~\ref{plotone} suggest a common source of enhanced 
scattering for many sight lines.
Given the current understanding of the LISM, and also on the basis of 
the results from our earlier work, we postulate that this enhanced 
scattering is probably caused by the shell of the Loop I Bubble.

\subsubsection{Modeling the Scattering due to the Loop I Shell}
\label{s:data-model}

Here we try to explain the observed discrepancies in decorrelation bandwidths
for nearby pulsars in the direction of Loop I in terms of enhanced scattering
from a combination of the Local Bubble and the Loop I Bubble. 
A full modeling of the problem involves the determination of the
parameters characterizing their size and location, and the distribution 
of scattering in and around them.  This was done by BGR98 for the case of
the Local Bubble alone.
However, in the present case, since we already have a model for the scattering 
from the Local Bubble and there are very good models for the physical
dimensions of Loop I, we do not really need to do such full-blown modeling.  
Instead, we take the following simpler approach.

First, we assume a scattering model for Loop I which is very similar to 
the multi-component model for the Local Bubble, i.e., 
an interior filled with a plasma of relatively low turbulence 
(the LOS-averaged strength of scattering, \avcn $\sim 10^{-4.2}$ \cnu)
and a dense, highly turbulent shell in which the scattering strength is 
many times larger than in the ambient ISM that lies beyond Loop I.  For 
this ambient medium, we assume \avcn$\sim 10^{-3.5}$ \cnu (see BGR98).

Second, as described in \S~\ref{s:data-sample}, the size and structure of 
Loop I are fairly well constrained by previous work (e.g. Berkhuijsen et 
al. 1971).
We simply adopt these values for characterizing the physical dimensions and 
location of the Loop I shell in our analysis. 

Third, we simplify the problem by assuming the thickness of the Loop I shell
(\dLI) to be small compared with the pulsar distances involved. 
The problem then essentially reduces to obtaining the best estimate of the
scattering measure due to the Loop I shell 
(\SMLI\, $\equiv \int _0 ^{\dLI} \cnLIsh (l) \ dl$, where \cnLIsh denotes 
the scattering strength inside the shell) that will minimize the 
discrepancies between \ndmeas and \ndpred. 
Note that the spherical shell geometry of Loop I implies that \SMLI\, should be a 
function of the LOS, or more precisely, of the angular distance of the LOS 
from the loop centre ($\Theta$). 
This is because the actual path length through the shell intercepted by the 
LOS varies with $\Theta$. 
The effect will be more prominent for sight lines that are close to being
``tangential.''
In the analysis that follows, we will constrain SM as defined for a LOS that 
is normal to the shell (i.e., along $\Theta =0$).

Fourth, instead of obtaining a solution that will give a best case minimum for
all the pulsars in Table~\ref{tableone}, we prefer to solve for the SM of the 
Loop I shell using only two pulsars PSRs J1744--1134 and J1456--6843. 
This is justified on the grounds that (a) for many pulsars in 
Table~\ref{tableone}, the distance uncertainties and the large errors in the 
estimates of the decorrelation bandwidth (due to RISS effects), make it 
less meaningful to do a global fit, and (b) for several pulsars at distances 
\ga 1 kpc, Loop I may {\it not} necessarily be the {\it only} source of 
enhanced scattering and a global fit for \SMLI\, could therefore be perturbed 
by this effect.

The choice of the above two pulsars is dictated by the following arguments.
Both pulsars are heavily scattered (Johnston et al. 1998) and have precise, 
independent distance estimates from parallax 
measurements --- $357^{+43}_{-35}$ pc for PSR J1744--1134 (Toscano et al. 1999) 
and $455^{+70}_{-56}$ pc for PSR J1456--6843 (Bailes et al. 1990).  
Interestingly, the estimate for PSR J1744--1134 is over twice the value of 
166 pc derived from the model of TC93.
The independent distance estimates allow the scattering geometry along 
these lines of sight to be fairly well constrained.  This is illustrated 
in Fig.~\ref{plottwo}, 
where the expected locations of the Local Bubble (\Db) and Loop I boundaries 
(\Dbone~ and \Dbtwo) in these directions are clearly marked.
The farther boundary of the Loop I shell (\Dbtwo) lies very close to
the mid-point 
to both objects, and is thus optimally placed for maximum contribution to 
the observed decorrelation bandwidth. 
Such a geometry will be consistent with the enhanced level of scattering 
inferred for these two pulsars.

Appendix A describes our method for computing the expected decorrelation
bandwidth (\ndpred) for a given distribution of the scattering material
along the LOS. 
We note that if homogeneously distributed scattering material 
(i.e., equation [A1], with the integral replaced by $\avcn D$, where
\avcn denotes the LOS-averaged strength of scattering) were to explain the 
observed scattering for the lines of sight of these pulsars, the 
implied level of scattering strength (\avcn =$10^{-2.8}$ \cnu towards
PSR J1744--1134, obtained from the value reported by Johnston et al. [1998],
after scaling for the new distance) would be 
several times larger than that expected for the typical ambient ISM. 
This seems quite unlikely.  
The LISM as relevant to the Local Bubble model of BGR98 can be treated 
as an inhomogeneous scattering medium, which can be best represented by
multiple components of different strengths of scattering (\cn) along
the LOS (see equation [A2], for example). This can be represented by
a simpler, modified version of equation (A2) (i.e., the right hand
side consisting only the first two integrals and the last integral 
with the lower limit, \Dbtwo + \dLI, replaced with \Db + \dLB).
If we were to attribute the entire excess scattering to 
material inside the Local Bubble shell (located at $\sim$30--60 pc in 
this direction; see Fig.~\ref{plottwo}), then \cnLBsh would have to be 
over an order of magnitude larger than that derived by BGR98;
alternatively, the shell would have to be much more extended in this 
direction (say, by an order of magnitude).
Again, either of these possibilities is rather unlikely, as it would require
a drastic change in the properties of the Local Bubble in this direction.
Hence the possibility that the Loop I shell is the source of the
enhanced scattering is most plausible. 

We now estimate the optimal value of \SMLI\, by modeling the scattering 
towards PSRs J1744--1134 and J1456--6843, taking into consideration the 
scattering due to the Local and Loop I bubbles, as well as that due 
to the distributed plasma (i.e., ambient ISM) beyond Loop I.
Our model also takes into consideration the characteristic fall-off with 
z-height for the ambient scattering plasma (scale height $\sim$500 pc; 
see TC93), and a Gaussian fall-off (scale height $\sim$125 pc; see BGR98 
for arguments) for the shell material.  The interiors of both bubbles are 
taken to be filled with weakly turbulent plasma ($\avcn \sim 10^{-4.2}$ \cnu).
The shell of the Local Bubble is taken to have a scattering measure 
(\SMLB) of 0.2 \smu (i.e., \avcn $\sim$ a few 100 times larger than that 
in the ambient medium).
The method for computing the expected decorrelation bandwidth (\ndpred) 
for such a geometry is described by equation (A2) in Appendix A. 
To determine \SMLI, we start with an initial value of 0.2 \smu, and vary this
from 0.02 to 2.0 \smu in steps of 0.01 \smu. 
Interestingly, the best agreement between the predicted \nd values and those
measured is obtained for \SMLI\, $\approx$ 0.29 \smu, a value quite 
comparable to that inferred for the Local Bubble shell.

Using the SM of the Loop I shell, as constrained by the above method, 
we compute new values for the \nd ratios.
These are shown by the filled star symbols ($\star$) in Fig.~\ref{plotone}.
It is quite remarkable that the above modeling successfully 
removes the discrepancies for pulsars out to a distance of a few 100 pc.
The $\epsA$ value for this model ($\approx~$0.55) is a significant 
improvement over that for the earlier model of the Local Bubble alone,
as listed in Table~\ref{tablethree}.  
Further, as can be seen from this table, there is a dramatic improvement 
in \epslow (i.e., for pulsars within 1 kpc), and a
much smaller one in \epshigh -- exactly as expected for the Local 
Bubble + Loop I Bubble model.

\subsection{Enhanced Scattering from the Sagittarius Spiral Arm}
\label{s:data-spiral}

Next we address the systematic downward trend for the ratio \ndratio
in Fig.~\ref{plotone} for pulsars beyond $\sim$1.2 kpc. 
A closer examination of these LOSs suggests material within the 
Sagittarius spiral arm to be the most plausible source of enhanced 
scattering for several of the pulsars. 
We show here that these results can be better explained by taking 
into account the enhanced level of scattering that can be expected 
due to this spiral arm as per the existing model of TC93.
With the arm parameters of the TC93 model and the distance estimates based 
on this model, the sight lines of six of our pulsars pass through this 
spiral arm.  
To model the effect of this, we have simply adopted the arm locations and 
density parameters of TC93. They consider a squared hyperbolic secant for 
the z-dependence and a Gaussian fall-off for the radial dependence for 
the arm density, with $ 300 \pm 100 $ pc for the scale height and 300 pc 
for the half width. 
With an arm density of 0.08 \cmc -- i.e., several times larger than the
typical \avne in the LISM, and a fluctuation parameter ($F$) of $ 6 ^{+5} _{-2} $, 
the implied level of scattering is some 2 orders of magnitude larger inside 
the arm (\cnsa$\sim$0.05 \cnu) than in the LISM.
Hence, substantial amounts of enhanced scattering are expected for these 
pulsars. 
On incorporating this, the discrepancies are further reduced 
(see the open circles in Fig.~\ref{plotone}), with a significant
improvement in overall agreement 
($\epsA \approx 0.22$) as well as for \epshigh (see Table~\ref{tablethree}). 
The best agreement is seen with the lower and upper limits allowed 
(by the model of TC93) for the values of the fluctuation parameter 
and the scale height, respectively.


\section{Results and Discussions}
\label{s:res}

We have shown that the combined effect of enhanced scattering from the Local 
Bubble shell, the Loop I Bubble shell and the Sagittarius arm goes a long way 
towards explaining the observed scattering properties of pulsars in
the general direction of Loop I.
Our new model consists of: (a) an ellipsoidal shell of SM$\sim$0.1--0.3 \smu 
to account for scattering due to the Local Bubble, (b) a spherical shell 
of SM$\sim$0.3 \smu to characterize the scattering due to Loop I, (c) 
the ambient ISM (\avcn $\sim 10^{-3.5}$ \cnu) in the inter-arm region, and 
(d) the Sagittarius spiral arm (\avcn $\sim$ 0.047 \cnu). 
We now discuss some of the implications of these results.

\subsection{Scattering from the Loop I Shell}
\label{s:res-loop1}

The agreement between the measurements and the predictions of our new
model is such that the scattering properties of most pulsars out to a 
distance of 1 kpc are successfully explained (\ndratio in the range
0.5--2.0 for 8 out of 10 objects).
The major outliers are PSRs J1713+0747 and J1730--2304, with residual 
discrepancies of $\sim 4$ and $\sim 3$ times respectively
between the predicted and measured values (see Fig. 1) that remain to 
be explained.  
These discrepancies are possibly due to RISS, or distance errors, or 
perhaps as yet un-modeled source(s) of enhanced scattering.  
For more distant pulsars (D \ga 1.2 kpc), although the discrepancies are better
accounted for, there still seems to be a systematic downward trend with
distance, which is accounted for by extra enhanced scattering due to the 
Sagittarius arm.
The results thus clearly substantiate the role of Loop I as a source of 
enhanced scattering in the LISM. 

We now address the various sources of uncertainties relevant to our
analysis.
These include:
(a) the uncertainties in size and geometry of (i) the Local Bubble,
and (ii) Loop I,
(b) the uncertainty in the (integrated) scattering strength of the
Local Bubble (\SMLB), 
(c) distance uncertainties, and 
(d) errors on the measurements of \nd.
First, we note that the Local Bubble geometry itself is not very well 
determined (see BGR98), and the expected location of the boundary is 
in the range $\sim$24--58 pc for PSR J1744--1134, and $\sim$44--74 pc 
towards PSR J1456--6843 (see Fig.~\ref{plotfour}). 
Taking this into consideration, \SMLI\, could be in the range 0.25--0.3 \smu.
However, the uncertainties in the angular size and location of Loop I itself 
are small (see Berkhuijsen et al. 1971), hence we do not expect
them to affect \SMLI\, appreciably. 
Turning to (b), the strength of scattering of the Local Bubble shell (\SMLB)
itself has a significant uncertainty : 0.11$<$\SMLB$<$0.28 \smu (see BGR98).
However, this seems to affect \SMLI\, only marginally (0.24--0.32 \smu).
The distances to PSRs J1744--1134 and J1456--6843 have $\sim$10\%
uncertainties, which translate in to a variation of 
0.28--0.33 \smu for \SMLI. 
As for the measurements of \nd, the dominant source of errors 
are due to their apparent variations caused by RISS effects 
on time scales from days to weeks.
However, these uncertainties are hard to estimate. 
Nevertheless, if we consider a typical case of a factor of two variation
in the measured values of \nd, \SMLI\, may lie in the range 0.15--0.57 \smu. 
Hence, the RISS-induced errors in measured decorrelation bandwidths are 
likely to be the dominant source of errors in the final estimate of the
strength of scattering of the Loop I shell material.

\subsection{Alternative scenario: Scattering from an Interaction Zone
between the Local Bubble and Loop I?}
\label{s:res-zone}

In the above analysis, we have attempted to understand the observations 
in terms of enhanced scattering caused by turbulent plasma within the 
shells of the two bubbles.
Although this two-bubble model seems to successfully explain the scintillation 
measurements for most nearby pulsars, there is an interesting
alternative that deserves consideration.
As described in \S~\ref{s:intro}, the work of EA95 and Egger (1998) 
suggest that the two bubbles are probably undergoing a collision
process; the observational evidence for a dense interaction
feature at $\sim$70 pc towards Sco-Cen (Fig. 4 of EA95) supports this.
Interestingly, the expected location of the closer boundary of Loop I 
(\Dbone) is either within the range of or near the Local Bubble
boundary (\Db) as constrained by BGR98 (see Fig.~\ref{plotfour}). 
In particular, for several lines-of-sight, \Dbone~lies in between 
the smaller and larger boundaries (\Dbsmall~and \Dblarge~
respectively) of the Local Bubble.
The possibility of enhanced scattering from an ``interaction zone'' 
between the two bubbles is therefore worthy of consideration.

To examine this possibility, we consider a model where the
measured level of enhanced scattering is {\it entirely} attributable to an 
``interaction wall'' or ``zone'' whose effective location is at the mean 
distance of the LB and Loop I boundaries. 
The two extreme possible geometries for the LB would thus imply two possible 
locations for the interaction zone (IZ): \Dbzone=0.5(\Dbsmall+\Dbone) 
and \Dbztwo=0.5(\Dblarge+\Dbone), which correspond to the cases of nearer 
and farther boundaries of the LB, respectively.
These are listed in column (5) of Table~\ref{tabletwo} (\dbz=\Dbz/D). 
In the discussion below, we will refer to these as the IZ-A and IZ-B 
models, respectively.
We also assume that the thickness of this zone (\diz) is comparable to 
that of the LB or LI shell, and also that it is much smaller than the 
pulsar distances.
Further, like the case of the Loop I shell, the path length of the sight 
line through a possible interaction zone region, and consequently its 
contribution to the scattering measure, will be a function of the angular 
distance of the LOS from the loop centre. 
In order to determine the scattering measure of the IZ (\SMIZ),
we performed a rather similar analysis to that described 
in \S~\ref{s:data-model}, in which PSRs J1744--1134 and J1456--6843 are 
used to estimate the value for \SMIZ. 
For the IZ-A model (\Dbz=\Dbzone), the best agreement between the measured
and predicted values for \nd is obtained for \SMIZ\, $\approx$1.12 \smu, which 
is $\sim$4 times larger than the value of SM for the Loop I shell, and
$\sim$4--10 times larger than that for the LB.
The required SM for the IZ-B case is somewhat smaller: $\sim$3 times the
value for \SMLI, and $\sim$3--9 times \SMLB.
This is mainly due to the relatively further location of the zone in this case.
Using these values for \SMIZ, we re-computed new values for \ndratio.
The results for the IZ-A model are shown in Fig.~\ref{plotthree}. 
The agreement is {\it not} as good as that achieved with the two-bubble
model, as indicated by somewhat larger values for the three $\epsilon$ parameters
that quantify the level of agreement (see Table~\ref{tablethree}).
On a closer comparison of Figs.~\ref{plotone} and \ref{plotthree}, it is obvious that
the agreement is somewhat poorer for several objects at D \la 1 kpc.
Nevertheless, the IZ model(s) may still be considered as a possible 
alternative to the two-bubble model.
However, unless there are well studied objects located in the Loop I interior,
it is not easy to distinguish conclusively between these two scenarios.

As a natural extension of the IZ model(s), we incorporate scattering
due to the Sagittarius spiral arm (as discussed in \S~\ref{s:data-spiral})
in order to improve upon the agreement for pulsars beyond 1 kpc with sight 
lines intercepting that arm. 
The results for the IZ-A+SA case are shown in Fig.~\ref{plotthree}. 
The overall agreement, as indicated by \epsall=0.27 and 0.25 respectively
for the IZ-A+SA and IZ-B+SA cases, is somewhat poorer than
that achieved with the two-bubble+SA scenario (\epsall=0.22, see 
Table~\ref{tablethree}).

\subsection{Discussion}
\label{s:res-disc}

\subsubsection{Implications of the Results}
\label{s:res-impli}
\begin{center}{\it (a) Interpretation of the Scintillation Data}\end{center}

From the results of this work it is quite evident that the structure of 
the LISM plays an important role in the interpretation of the scintillation 
data.
The outer shells of the Local Bubble and Loop I may very well be the dominant 
sources of scattering towards many directions in the LISM, especially
at higher Galactic latitudes. 
Interestingly, independent evidence in favor of similar close-by scattering
screens ($\sim$25--250 pc) comes from 
recent observations of cm-wave ISS and intra-day variability (IDV) of 
quasars (e.g., Dennett-Thorpe \& de Bruyn 2000; Rickett 2000).
In particular, the screen location of $\sim$25 pc and strength of scattering,
\cn $\approx$0.2 \cnu, inferred by Dennett-Thorpe \& de Bruyn (2000) towards
the IDV quasar J1819+3845 is in very good agreement with the location of the
Local Bubble boundary ($\sim$25--50 pc) and the scattering strength (\SMLB) 
expected in this direction. 
Further, from the work of Hjellming \& Narayan (1986), the bulk of the
scintillation (RISS) of the radio source PKS 1741--038 at 1.49 GHz 
is caused by a single screen located at $\sim$140 pc from the Sun.
Interestingly, the LOS to this object ($l=21^{\circ}.6, b=13^{\circ}.1$)
lies very close to the ``inner ridge feature'' of the NPS (beginning at
$l\approx22^{\circ}, b=14^{\circ}$).
Also, Hjellming \& Narayan infer that there is significant excess scattering 
(\cn $\sim 10^{-1.5}$ \cnu) along this LOS. 
Recently, Lazio et al. (2000) report multi-epoch VLBI observations of this 
object as it underwent an extreme scattering event (ESE); the inferred 
level of scattering for the ESE lens is orders of magnitude larger than 
that in the ambient medium.
Given the close proximity of the object to Loop I, it is quite probable
that the structure(s) that caused the ESE are located at the Loop I shell.
Similarly, if an interaction wall between the two bubbles exists, it
could potentially act as a thin scattering screen for nearby pulsars with 
sight lines within $270^{\circ}<l<30^{\circ}$ and $-40^{\circ}<b<+80^{\circ}$.
Such close-by scattering screens may also explain the shorter than 
expected timescales seen with the slow intensity variations of pulsars
and the low
frequency variables (Gupta, Rickett \& Coles 1993; Spangler et al. 1993).
In addition, the enhanced scattering from such bubble shells may be responsible
for some of the unusual scattering effects.
In this context, it is interesting to note that the location of the scatterer that 
caused the multiple imaging event of PSR B1133+16 in the Ooty data (Gupta, Bhat \& Rao 1999)
was found to match well with the expected location of the Local Bubble shell 
(\Db~$\approx$~0.77\,D) in this direction. 
Similarly, it has also been suggested that the the structures that cause
extreme scattering events in the radio light curves of some quasars
are probably associated with the Galactic loops (Fiedler et al. 1994).
Recent work by Toscano et al. (1999) and Chatterjee et al. (2000) provides some
evidence for enhanced scattering in the third galactic quadrant probably 
associated with an interface region between the Local Bubble and the 
GSH 238+00+09 super-bubble.
All of the above arguments clearly support the view that the structure
of the LISM needs to be taken into account for a proper interpretation of 
dispersion and scintillation data.

\begin{center}{\it (b) On the Nature of the Bubble Shells}\end{center}

In our earlier work (BGR98), we argued that the elongated ellipsoidal shell 
of turbulence derived from the Ooty data is possibly associated with the 
Local Bubble.
The analysis described in \S~\ref{s:data-loop1} presents quite convincing 
evidence for an outer turbulence shell for Loop I also (the existence of a 
turbulent interaction zone is an alternative possibility, though). 
Thus, from observational data, it appears that interstellar bubbles in 
general may have turbulent outer shells.

On the theoretical front, a number of authors have dealt with the evolution 
of bubbles in the ISM (Ikeuchi 1998 and references therein); in particular, 
special attention has been paid to the major local features such as the Local 
and Loop I Bubbles.
Interestingly, it turns out that turbulent outer shells can indeed be
expected for interstellar bubbles. 
Specifically, the recent model of Breitschwerdt, Freyberg \& Egger (2000) 
predicts turbulence to exist at the interface region of the Local and Loop I
Bubbles as a natural consequence of the hydro-magnetic 
Rayleigh-Taylor instability caused by the interaction of the two bubbles.
A similar mechanism may also apply for the Loop I shell. 
In an earlier work, Breitschwerdt \& Kahn (1989; see also Kahn \&
Breitschwerdt 1988) showed that turbulent shells could also be 
expected for stellar-wind--blown bubbles as a consequence of 
acoustic instability.
So, in general, the shells around bubbles may be expected to be 
turbulent.  There are no explicit predictions available for the 
level of turbulent intensity expected at the shell or in interaction 
zone regions, or for a relation connecting the turbulent intensity to 
the physical parameters governing the process(es).
Nevertheless, the basic results from our investigations using pulsar 
data appear to be in accordance with some of the theoretical models of 
bubble evolution.  

\subsubsection{Estimation of the Shell Density}
\label{s:res-nesh}

The strength of scattering we derived for the outer shell of Loop I 
is substantially higher ($\sim$100--500 times depending on the shell
thickness) than that measured towards
typical LOSs in the LISM.
The regions of enhanced electron density (\nele) are also thought to 
be plausible sites for enhanced scattering (e.g. the Gum Nebula), 
though the exact relationship between these two is not clearly known. 
If we consider \cn~to be uniform within the shell or the interaction
zone regions (i.e., \SMLI.$\equiv$\cnLIsh \dLI, \SMIZ\,$\equiv$\cnIZ
\diz, where \dLI and \diz represent the path lengths through the 
shell and interaction zone regions respectively), 
and assume a simple relation ($\cn \propto \nelesq$), then an indirect 
estimate of electron density in the shell (\nesh) is possible for a 
given shell thickness.
Considering the Loop I shell (SM$\sim$0.3 \smu), for a shell thickness $\sim$1--10 pc,
this yields a density $\sim$10--30 times larger than the ambient ISM value.
Somewhat higher values ($\sim$20--50\avne) result 
for the interaction zone models described in \S~\ref{s:res-zone}.
Interestingly, a similar level of density enhancement is seen for the neutral 
gas at the annular interface region of the two bubbles (EA95; Egger 1998).
For the above densities, the effects on dispersion due to the shell region 
will be significant only for pulsars within a distance of a few 100 pc. 
Taking this into consideration would, however, result in a {\it much lower} 
ambient density (\avne $\approx$0.007 \cmc) towards PSR J1744--1134, 
the disk pulsar with the lowest known \nele. 
For a similar density enhancement ($\sim$10 times) for the shell region 
towards PSR J1456--6843 (the only pulsar in quadrant 4 with a measured 
parallax), consideration of shell dispersion would result in an 
ambient density of $\approx$ 0.016 \cmc\, for the two-bubble model, and 
$\approx$ 0.013 \cmc\, for the interaction zone model.
These values are comparable to \avne measured towards several objects 
in quadrant 1 with known parallaxes (see Toscano et al. 1999).
Therefore, it is quite plausible that the Loop I shell (or the interaction 
zone) contributes significantly to the dispersion of nearby pulsars in 
quadrant 4.  

\subsubsection{Electron Densities towards PSRs J1744--1134 and J1456--6843:
Evidence for a Dense Wall?}
\label{s:res-nelism}

It is interesting to note that for the two pulsars -- PSRs J1456--6843 and 
J1744--1134 -- with independent distance estimates, the mean
electron densities
along the sight lines differ significantly: \avne towards the former
is almost twice that for the latter, despite their comparable distances and
z-heights (67 and 57 pc respectively).
A closer look at their LOSs {\it vis-a-vis} the geometries of the 
Local and Loop I Bubbles (Figs.~\ref{plottwo} and~\ref{plotfour}) indicates the following:
(1) the Local Bubble cavity extends out almost twice as far towards 
PSR J1456--6843 as towards PSR J1744--1134 ($\sim$44--74 pc and $\sim$24--58 pc
respectively, for the geometry in Fig.~\ref{plotfour}),
(2) the interior of Loop I covers a much longer section of the 
LOS towards PSR J1456--6843 (\Dint\,$\equiv$\,\Dbtwo--\Dbone\, $\approx$\, 233 pc
for J1456--6843, compared to 157 pc for PSR J1744--1134; see Fig.~\ref{plottwo}), 
and (3) the extent of the ambient medium beyond Loop I is not much larger 
along the sight line towards PSR J1456--6843
($\sim$192 pc compared to $\sim$159 pc for PSR J1744--1134).
All of these make it harder to explain the larger \avne 
towards PSR J1456--6843.

There appears to be two plausible explanations for the above:
(a) the ambient ISM in quadrant 4 is simply much denser 
than that in quadrant 1. 
However, in the absence of other pulsars in quadrant 4 with 
measured parallax, there seems to be no independent way to confirm this.
(b) The existence of a dense interface
region between the Local and Loop I Bubbles, as described in 
\S~\ref{s:res-zone}.
On a closer inspection of the sight lines of the two pulsars, 
case (b) seems the more likely.
The LOSs towards PSR J1744--1134 intercepts the ``annular \HI ring,''
as recognized by EA95 (see also Egger 1998),
whereas the LOS toward PSR J1456--6843 lies well within the 
``interaction zone region'' confined by this ring-like feature.
If the interaction zone region is filled with a denser, highly 
turbulent plasma, then it could potentially account for much 
of the dispersion towards PSR J1456--6843.
If we assume an ambient density of $\approx$ 0.01 \cmc (i.e., 
a value comparable to that in quadrant 1), the interaction 
zone density (\neiz) will 
have to be $\sim$40 times larger in order to account for the 
excess dispersion. 
We note that this is quite comparable to the value $\sim$20--50\avne 
derived from the SM estimate as constrained by our modeling.
The consistency of the density estimates from the two independent
methods can be argued, in some sense, to favor the existence of 
a dense interaction wall. 

Measurements of parallax for other nearby pulsars in quadrant
4 will help to test for the existence of a wall.
The most promising candidates for this seem to be low-DM (say, \la 20 \dmu)
objects with LOSs intercepting the interaction zone region. 
For instance, PSR J1751--4657 would be located much farther ($\sim$1765 pc)
than its TC93 distance ($\sim$1080 pc), if the ambient \nele in this quadrant 
is comparable to that in quadrant 1 ($\sim$0.01 \cmc).
PSR J1730--2304 (close-by to PSR J1744--1134) is another interesting test case, whose distance will only
be slightly more ($\approx$555 pc) if a denser zone is present, while 
it will be located much farther away ($\sim$1 kpc) if the entire sight
line was uniformly filled at \nele$\sim$0.01 \cmc.
Similarly, the distance of PSR J1455--3330 will be $\sim$950 pc (compared
to TC93 value of $\sim$740 pc) if we consider a dense zone and a low-density
($\approx$0.01 \cmc) ambient medium. 
While our arguments are based on a fairly simple picture, it is amply clear
that measurements of a few interesting test cases will be valuable for 
understanding the LISM in this quadrant. 

\subsubsection{Pulsars and the LISM: Further Prospects}
\label{s:res-pros}

From the work presented here and related work in the recent
past (see \S~\ref{s:res-impli}), it is clear that pulsars can be used 
quite effectively to
probe large-scale features in the LISM.  Conversely, it is also obvious
that detailed interpretation of pulsar dispersion and scintillation data
will need to take the structure of the LISM into account.  
There is growing observational evidence that the Local Bubble is surrounded 
by numerous bubbles of similar properties, many of which are likely to be 
filled with hot X-ray--emitting gas.
While this work has concentrated on Loop I, our most prominent  
such neighbor, other nearby examples include the Eridanus bubble, the 
Gum Nebula and possibly radio Loops II and III.  
Further, there are also observations indicating possible interactions of
some of these with the Local Bubble.  In this paper, we have considered
the possibility of an interaction region between the Local and Loop I Bubbles.
Toscano et al. (1999) and Chatterjee et al. (2000) have provided some evidence
for interaction of the Local Bubble with the GSH 238+00+09 super-bubble 
in galactic quadrant 3.
The closeness between the location of the near side of the Eridanus Bubble 
(159$\pm$16 pc; Guo et al. 1995) and the expected location of the Local 
Bubble shell in this direction ($\approx$130 pc) 
suggests a possible interaction between these two bubbles.
The Vela supernova remnant has been shown to be embedded in a hot 
bubble confined by the shell of the Gum Nebula 
(Aschenbach, Egger \& Tr\"umper 1995).
With the recently revised estimate for its distance (250$\pm$30 pc by 
Cha, Sembach \& Danks 1999), the near side of the ``Gum-Bubble'' is 
quite close to the Local Bubble boundary ($\approx$100 pc); this is 
also supported by the observation that 
the absorbing column density towards the Vela SNR is only 
about $10^{20}$ \cms.
An interaction between these two bubbles therefore seems quite likely.

Of the 163 pulsars known within $\sim$2 kpc of the Sun, 
scintillation data are presently available for only 73,
and independent distance estimates for only 12.
Clearly, a lot more needs to be done here.
The new generation of large telescopes, such as the GMRT 
and the GBT, could potentially extend the available scintillation 
data for nearby weak pulsars.
For many such cases, measuring pulse-broadening times at low frequencies
(\la 100 MHz) may prove to be a more viable technique than decorrelation 
bandwidth measurements.
Observations of sources along well-chosen lines of sight where the scattering 
is dominated by the shells of these bubbles may also result in more detections
of unusual scattering effects, such as multiple imaging and extreme 
scattering events. 
Further, ISS studies of radio sources through cm-wave ISS and IDV
may also provide useful insights into the nature of the LISM at 
higher Galactic latitudes.
Many such observations hold great promise for extending the rather
simplistic models derived from the present investigations into a 
more realistic picture of the LISM.


\section{Conclusions}
\label{s:conc}

We have investigated the distribution of scattering plasma in the LISM
in the general direction of the Loop I Bubble by combining
recent pulsar scintillation measurements.
Many pulsars within $\sim$2 kpc, and located towards 
$270^{\circ} < l < 30^{\circ}$ and $-40^{\circ} < b < +80^{\circ}$, show 
enhanced levels of scattering, detected as a significantly reduced
value of the ratio of the measured to expected decorrelation bandwidths.
The discrepancies cannot be explained by the Local Bubble model
(BGR98) alone, and
we interpret them as being due to enhanced scattering associated with the 
shell of the Loop I Bubble and the Sagittarius spiral arm.
Using data from two heavily scattered pulsars with precisely known distances, 
PSRs J1744--1134 and J1456--6843, we have placed useful 
constraints on the scattering strength associated with the Loop I shell. 
The scattering measure inferred for the plasma inside 
this shell is $\sim$0.3 \smu, 
quite similar to that for the Local Bubble shell.
Assuming a shell thickness $\sim$1--10 pc, this implies an average 
turbulence level $\sim$100--500 times larger than that in the 
ambient ISM.
Adopting this, our earlier model of the LISM is extended  
by incorporating an explicit scattering component for plasma in and around 
Loop I, in addition to that due to the Local Bubble.  
This two-bubble model successfully explains many of the observed 
decorrelation bandwidth discrepancies for nearby pulsars.
The alternative possibility of enhanced scattering from an interaction zone 
between the Local Bubble and Loop I is also considered. 
However, the level of turbulence in the interaction zone would have to
be several times larger than that for the shells in the two-bubble
model in order to explain 
the observations.  Even then, the final agreement with observations for
the interaction zone model is not as good as that for the two-bubble model.
Assuming a simple relation between the scattering strength 
and the free electron density (\cn$\propto$\nelesq) yields 
a shell density that is $\sim$10--30 times larger than the ambient 
value, and a somewhat higher density ($\sim$20--50\avne) for the 
interaction zone model.
For several low-latitude pulsars at D$\sim$1--2 kpc ($|z|$\la 300 pc), we find
that the observed scattering discrepancies are consistent with additional 
enhanced scattering from the Sagittarius spiral arm.

We have discussed implications of our results for the interpretation
of scintillation data as well as for the structure of the LISM.
In the light of our results and those from several other recent works,
we conclude that the structure of the LISM needs to be considered
in the interpretation of pulsar dispersion and scintillation data, and 
may also be relevant for observations of cm-wave ISS.
Further, the general picture that emerges from the investigations
in support of turbulent outer shells and/or interface regions of
interstellar bubbles appears to be in qualitative agreement with the
expectations of the models that describe the evolution of bubbles
in the ISM. 

\noindent
{\it Acknowledgments:} We thank C. Salter for several fruitful
discussions and valuable comments on the earlier versions of this
paper which helped us to improve upon the content. We also thank
P. B. Preethi for help with the analysis software during early 
stages of this work.  

\appendix
\section{Decorrelation bandwidth for an inhomogeneous, multi-component 
scattering medium}
\label{s:append}

\begin{figure}
\hskip 0.5in
\psfig{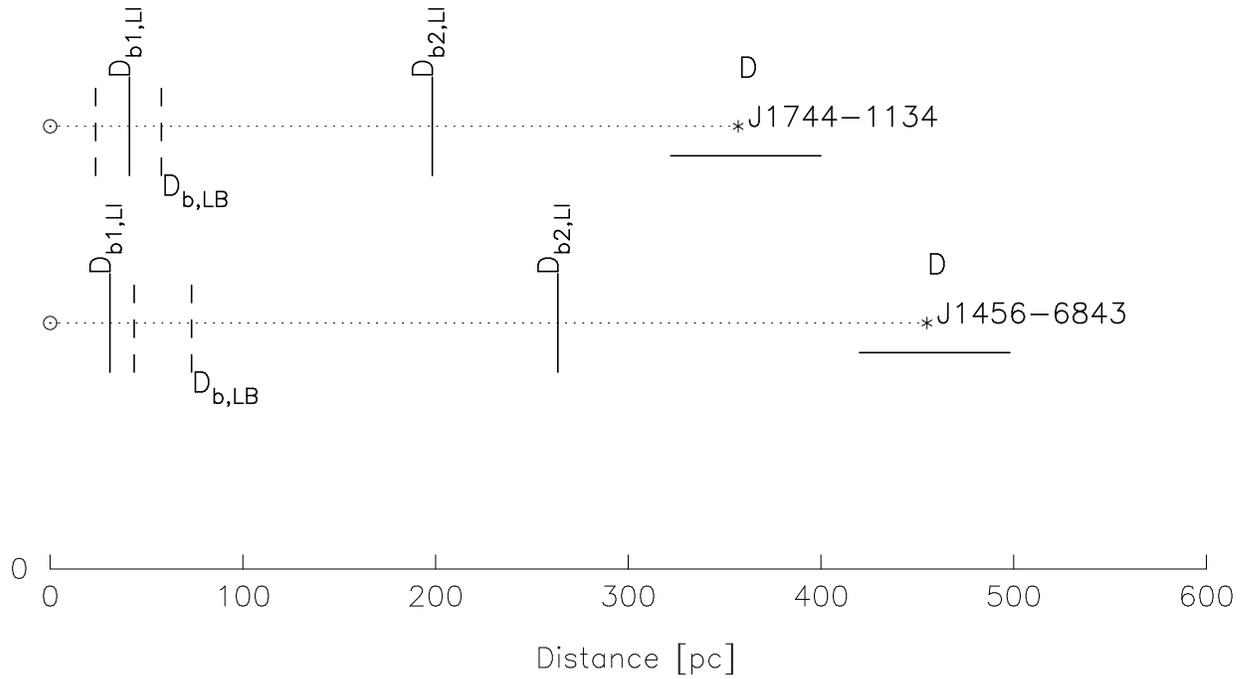}
\caption[]{Locations of the Local Bubble and Loop I shells along the sight 
lines towards PSRs J1744--134 and J1456--6843. 
The solid lines (\Dbone~and~\Dbtwo) indicate the Loop I boundaries, and 
the dashed lines are the positions of the Local Bubble boundary (\Db) 
that correspond to the two envelopes as shown in Fig.~\ref{plotfour}. 
The horizontal bar near the asterisk symbol ($\ast$) indicates the uncertainty 
in the distance estimate of the pulsar.}
\label{plottwo}
\end{figure}

\begin{figure}
\hskip 0.5in
\psfig{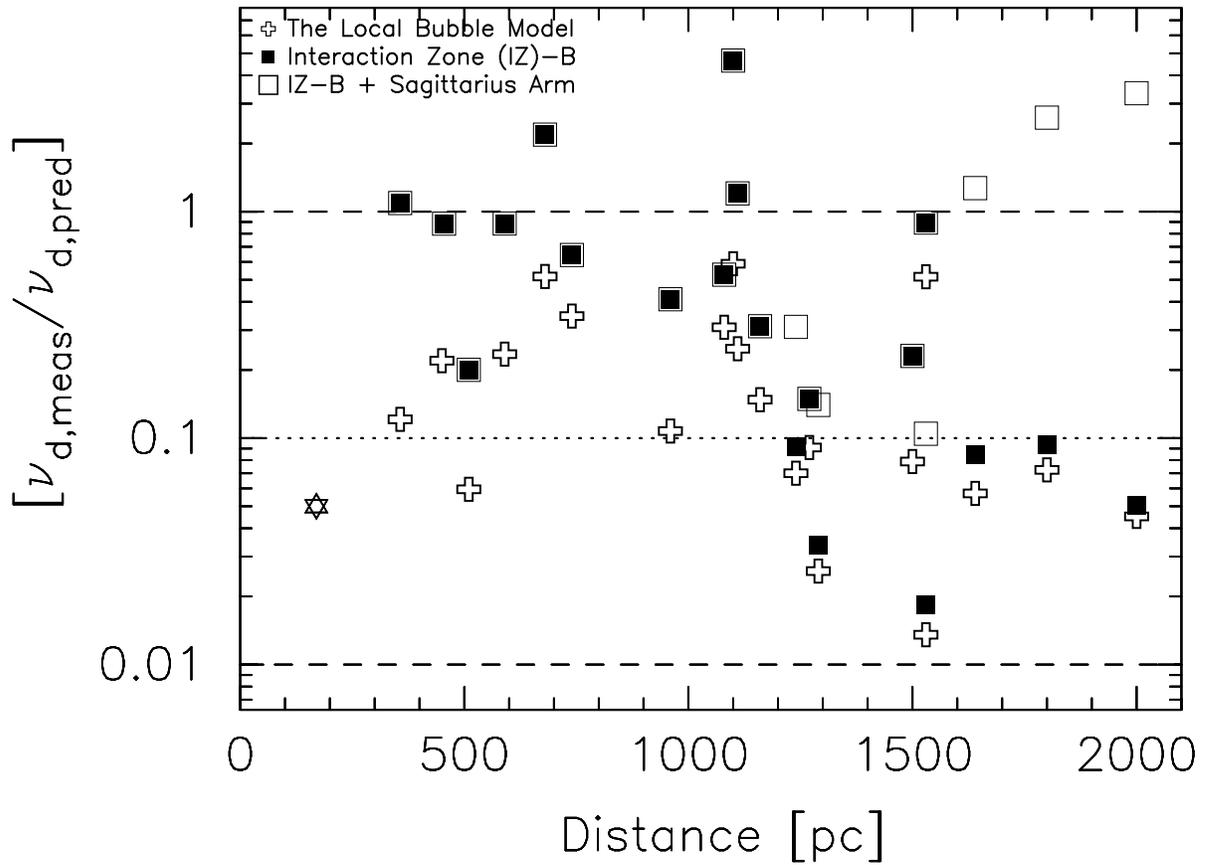}
\caption[]{Same as in Fig.~\ref{plotone} for the model where the enhanced 
scattering is due to a possible interaction zone between the Local Bubble 
and the Loop I Bubble. The results are for \Dbz=0.5(\Dbsmall+\Dbone), i.e., 
for the IZ-A+SA model as described in Table~\ref{tabletwo} and 
\S~\ref{s:res-zone}.}
\label{plotthree}
\end{figure}

\begin{figure}
\hskip 0.5in
\psfig{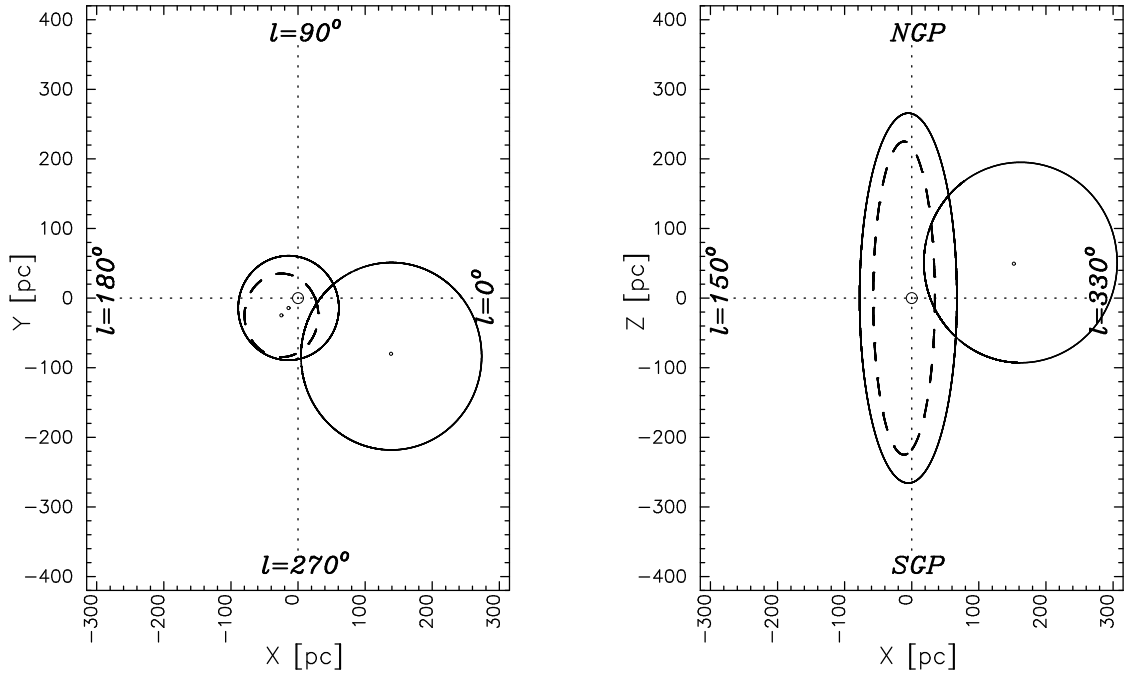}
\caption[]{Geometry of the Local Bubble and the Loop I Bubble; panel (a) is 
the section in the Galactic plane, and panel (b) is the section along a 
plane perpendicular to the Galactic plane and passing through the north 
and the south Galactic poles, as well as through the centre of of Loop I. 
The dashed and solid curves of the elongated cavity correspond to the 
two different geometries for the local scattering structure as derived 
from Ooty scintillation data (cf. BGR98).}
\label{plotfour}
\end{figure}

Here we describe the method adopted for computing the decorrelation 
bandwidth (\ndpred).
Fig.~\ref{plottwo} depicts typical scattering geometries relevant in our 
analysis: \Db is the expected location of the Local Bubble boundary,
\Dbone~and \Dbtwo~denote the nearer and farther boundaries, respectively, 
of the Loop I Bubble. D is the pulsar distance.

For a homogeneous scattering medium, the decorrelation bandwidth (\nd) 
is given by (CWB85)

\begin{equation}
\nd ~ = ~ { 1 \over D } ~ \left( \Aalpha \fobsalpha \right) 
^{\left( { 2 \over \alpha - 2 } \right)} 
~ \left( \int _0 ^D C_n^2 (z) ~ dz \right) ^ { 2 \over 2 - \alpha }
\end{equation}

\noindent
where \Aalpha is a model dependent constant, \fobs denotes the frequency 
of observation, and $\alpha$ is the slope of the electron density wavenumber 
spectrum. In this scheme, the observer is at $z=0$ and the pulsar at $z=D$. 
If we adopt the canonical value of 11/3 for $\alpha$ 
(i.e., Kolmogorov-like spectrum), $ \Aalpha= 2 \times 10^{-6} $.

The LISM model corresponding to Fig.~\ref{plottwo} can be treated as an 
inhomogeneous scattering medium, with multiple components of different 
strengths of scattering (\cn, also called turbulent intensity or 
scattering strength), 
located at different points along the line of sight. 
The observable \nd is determined by the path length differences of scattered 
rays; therefore, contributions to it from \cnz need to be appropriately 
weighted in such a way that scattering regions near the source or the 
observer produce smaller path-length differences than those that are 
mid-way (CWB85). 
Hence, for the case of our interest, equation (A1) can be 
rewritten as 

\begin{eqnarray}
\nd = 
{ 1 \over D }~ \left( \Aalpha \fobsalpha \right) 
^{\left( { 2 \over \alpha - 2 } \right)} ~
\left\{ \int _0 ^{\Db} \wz \cnLBint (z) dz +
\int _{\Db} ^{\Db + \dLB} \wz \cnLBsh (z) dz + \right. \nonumber \\
\int _{\Db + \dLB } ^{\Dbone } \wz \cnism (z) dz + 
\int _{\Dbone } ^{\Dbone + \dLI } \wz \cnLIsh (z) dz + \nonumber \\
\int _{\Dbone + \dLI } ^{\Dbtwo } \wz \cnLIint (z) dz +
\int _{\Dbtwo } ^{\Dbtwo + \dLI } \wz \cnLIsh (z) dz + \nonumber \\
\left. \int _{\Dbtwo + \dLI } ^D \wz \cnism (z) dz 
\right\} ^{\left( { 2 \over 2 - \alpha } \right)}
\end{eqnarray}

\noindent
where \dLB and \dLI are the thickness of Local Bubble and Loop I shells, 
respectively. 
The turbulent intensity in the Local Bubble interior and in its
shell are represented by \cnLBint and \cnLBsh, respectively, and 
\cnLIint and \cnLIsh are the equivalent quantities for the Loop I Bubble. 
The scattering plasma in the ambient ISM is characterized by \cnism.
The symbol \wz is the `weighting function' for \cnz, and is given by 

\begin{equation}
\wz ~ = ~ { z \over D } ~ \left( 1 - {z \over D} \right)
\end{equation}

This simple function is symmetric with respect to the mid-point between 
observer and pulsar, which means an inherent ambiguity is involved in the 
interpretation of the underlying scattering geometry. 
For instance, in the simple case of a thin screen placed between us and 
a pulsar, a screen closer to pulsar (say, at $z={3 \over 4}D$) will be 
equivalent to the one placed at $z={1 \over 4}D$ from us.
In the case of an extended, inhomogeneous medium (as in Fig.~\ref{plottwo}), the 
`inverse geometry' is indistinguishable from the actual geometry.


\bigskip
\noindent{\large\bf References:}
\bigskip
 

\noindent
Aschenbach, B., Egger, R. J., Trumper, J., 1995, Nature, 373, 587 \\
Bailes, M., Manchester, R. N., Kesteven, M. J., Norris, R. P., \&
Reynolds, J. E. 1990, Nature, 343, 240 \\
Berkhuijsen, E., Haslam, C. G. T., Salter, C., 1971, A\&A, 14, 252 \\
Bhat, N. D. R., Gupta, Y., Rao, A. P., 1998, ApJ, 500, 262 (BGR98) \\
Bhat, N. D. R., Gupta, Y., Rao, A. P., 1999a, ApJ, 514, 249 \\
Bhat, N. D. R., Rao, A. P., Gupta, Y., 1999b, ApJS, 121, 483 \\
Breitschwerdt, D., Freyberg, M. J., Egger, R, J., 2000, A\&A, 361, 303 \\
Breitschwerdt, D., Freyberg, M. J., Tr\"umper, J., 1998,
{\it Lecture Notes in Physics}, 506 \\
Breitschwerdt, D. \& Kahn, F. D., 1989, MNRAS, 235, 1011 \\
Centurion, M., \& Vladilo, G., 1991, ApJ, 372, 494 \\
Cha, A. N., Sembach, K. R., Danks, A. C., 1999, ApJ, 515, L25 \\
Chatterjee, S., Cordes, J. M., Lazio, T. J. W., Goss, W. M., Fomalont, E. B.,
Benson, J. M., 2000, ApJ, submitted. \\
Cordes, J. M., 1986, ApJ, 311, 183 \\
Cordes, J. M., Weisberg, J. M., \& Frail, D. A., et al. 1991,
Nature, 354, 121 \\
Cordes, J. M. \& Rickett, B. J. 1998, ApJ, 507, 846 \\
Cordes, J. M., Weisberg, J. M. and Boriakoff, V. 1985, ApJ, 288, 221, (CWB85)
\\
Cordes, J. M., Wolszczan, A., Dewey, R. J., Blaskiewicz, M., Stinebring, D.
R., 1990, ApJ, 349, 245 \\
Cox, D. P., \& Reynolds, R. J. 1987, ARA\&A, 25, 303 \\
Dennett-Thorpe, J. \& de Bruyn, A. G. 2000, ApJ, 529, L65 \\
Egger, R. J., 1993, Ph.D. Thesis, TU Muenchen \\
Egger, R. J. 1998, Lecture Notes in Physics, 506, 287 \\
Egger, R. J., \& Aschenbach, B. 1995, A\&A 294, L25, (EA95) \\
Fiedler, R. L., Dennison, B., Johnston, K. J., Waltman, E. B. and Simon, R.
S. 1994, ApJ, 430, 581 \\
Frisch, P. C. 1981, Nature, 293, 377 \\
Frisch, P. C. 1996, Space Sci. Rev. 78, 213 \\
Frisch, P. C. 1998, Lecture Notes in Physics, 506, 269 \\
Gothoskar, P. \& Gupta, Y. 2000, ApJ, 531, 345 \\
Guo, Z., Burrows, D. N., Sanders, W. T., Snowden, S. L., Penprase, B. E.,
1995, ApJ, 453, 256 \\
Gupta, Y., Bhat, N. D. R. \& Rao, A. P., 1999, ApJ, 520, 173 \\
Gupta, Y., Rickett, B. J. and Coles, W. A. 1993, ApJ, 403, 183 \\
Gupta, Y., Rickett, B. J., \& Lyne, A. G., 1994, MNRAS, 269, 1035 \\
Gwinn, C. R., Bartel, N., Cordes, J. M., 1993, ApJ, 410, 673 \\
Hajivassiliou, C. A. 1992, Nature, 355, 232 \\
Haslam, C. G. T., Khan, F. D. \& Meaburn, J., 1971, A\&A, 12, 388 \\
Haslam, C. G. T., Large, M. I. \& Quigley, M. J. S., 1964, MNRAS, 127, 273
\\
Heiles, C., 1998, ApJ, 498, 689 \\
Hjellming, R. M. \& Narayan, R., 1986, ApJ, 310, 768 \\
Ikeuchi, S., 1998, Lecture Notes in Physics, 506, 399 \\
Johnston, S., Nicastro, L., \& Koribalski, B. 1998, MNRAS, 297, 108 \\
Kahn, F. D. \& Breitschwerdt, D., 1988, MNRAS, 242, 209 \\
A\&A,
287, 470 \\
Lallement, R. 1998, Lecture Notes in Physics, 506, 19 \\
Lazio, T. J. W., Fey, A. L., Dennison, B., et al., 2000, ApJ, 534, 706 \\
Milne, D. K., 1968, Australian J. Phys., 21, 201 \\
Moran, J. M., Greene, B., Rodriguez, L. F. \& Backer, D. C., 1990,
ApJ, 348, 147 \\
Nishikida, K., 1999, Ph.D. Thesis, The Pennsylvania State University \\
Phillips, J. A. \& Clegg, A. W. 1992, Nature, 360, 137 \\
Rickard, J. J., \& Cronyn, W. M. 1979, ApJ, 228, 755 \\
Rickett, B. J., 1990, ARA\&A, 28, 561 \\
Rickett, B. J., 2000, Proceedings of IAU 182 Coll., Kluwer Academic
Publishers \\
Salter, C. 1983, BASI, 11, 1 \\
Spangler, S. R., Eastman, W. A., Gregorini, L., et al. 1993, A\&A, 267, 213 \\
Stinebring, D. R., Faison, M. D. \& McKinnon, M. M., 1996, ApJ, 460, 460 \\
Taylor, J. H., \& Cordes, J. M. 1993, ApJ, 411, 674, (TC93) \\
Toscano, M., Britton, M. C., Manchester, R. N., et al. 1999, ApJ, 523, L171 \\
Tr\"umper, J., Adv. Space Res. 1983, 4, 241 \\
Warwick, R. S., Barber, C. R., Hodkin, S. T., Pye, J. P., 1993,
MNRAS, 262, 28 \\
Yoshioka, S., \& Ikeuchi, S. 1990, ApJ, 360, 352 \\
\begin{table}
\caption{Pulsar Sample: The scintillation data}
\begin{center}
\begin{tabular}{lrrrllc}
\hline
\hline
            &      &        &     &       &     &    \\
PSR         & $l$  & $b$    & D   &\ndmeas&\fobs&Ref.\\
            &[deg] &[deg]   &[pc] & [MHz] &[MHz]&    \\
(1)         &(2)   & (3)    & (4) &(5)    &(6)  &(7) \\
\hline
\hline
J1057$-$5226& 286.0& $  6.6$& 1530& 2.52  & 660 & 1\\   
J1430$-$6623& 312.7& $ -5.4$& 1800& 0.28  & 660 & 1\\
J1455$-$3330& 330.7& $ 22.6$&  740& 1.37  & 436 & 1\\
J1456$-$6843$^{\ddagger}$                 
	        & 313.9& $ -8.5$&  455& 1.4   & 436 & 1\\
J1537$+$1155&  19.8& $ 48.3$&  680& 1.53  &  436& 1\\
J1543$-$0620&   0.6& $ 36.6$& 1160& 0.111 & 327 & 2\\
J1544$-$5308& 327.3& $  1.3$& 1290& 6.8   & 1520& 1\\
J1559$-$4438& 334.5& $  6.4$& 2000& 0.16  & 660 & 1\\
J1603$-$7202& 316.6& $-14.5$& 1640& 0.36  & 660 & 1\\
J1605$-$5257& 329.7& $ -0.5$& 1240& 20.2  & 1520& 1\\
J1607$-$0032&  10.7& $ 35.5$&  590& 0.379 & 327 & 2\\
J1614$+$0737&  20.6& $ 38.2$& 1500& 0.16  & 436 & 1\\
J1709$-$1640&   5.8& $ 13.7$& 1270& 0.040 & 1000& 4\\
J1713$+$0747&  28.8& $ 25.2$& 1100& 1.45  &  436& 1\\
J1730$-$2304&   3.1& $  6.0$&  510& 0.17  & 327 & 3\\
J1744$-$1134$^{\dagger}$                  
	        &  14.8& $  9.2$&  357& 1.34  & 436 & 1\\
J1751$-$4657& 345.0& $-10.2$& 1080& 0.165 & 327 & 2\\
J1752$-$2806&   1.5& $ -1.0$& 1530& 0.003 & 1000& 4\\
J1848$-$1952&  14.8& $ -8.3$&  960& 0.23  &  436& 1\\
J2053$-$7200& 321.9& $-35.0$& 1110& 0.55  &  436& 1\\
\hline
\end{tabular}
\end{center}
References: 
{(1) Johnston, Nicastro \& Koribalski (1998), 
(2) Bhat, Rao \& Gupta  (1999b), 
(3) Gupta \& Gothoskar (2000), (4) Cordes (1986)} \\
$^{\ddagger}$ Interferometric parallax distance from Bailes et al. (1990) \\
$^{\dagger}$ Timing parallax distance from Toscano et al. (1999) \\
\label{tableone}
\end{table}


\begin{center}
{ }
\end{center}

\begin{table}
\caption{Locations of the LB and LI boundaries and the IZ }
\begin{center}
\begin{tabular}{lcccc}
\hline
\hline
            &            &     &     &             \\
PSR         &\db$^{\ast}$&\dbone$^{\star}$
                               &\dbtwo$^{\star}$
                                     &\dbz         \\
(1)         & (2)        & (3) & (4) & (5)         \\
\hline
\hline
J1057$-$5226&0.04$-$0.05&0.03&  0.14& 	0.03$-$0.04   \\
J1430$-$6623&0.03$-$0.04&0.02&  0.15& 	0.02$-$0.03   \\
J1455$-$3330&0.05$-$0.10&0.04&  0.42& 	0.04$-$0.07   \\
J1456$-$6843&0.10$-$0.16&0.07&  0.57& 	0.08$-$0.12   \\
J1537$+$1155&0.05$-$0.12&0.07&  0.24& 	0.06$-$0.10   \\
J1543$-$0620&0.03$-$0.06&0.03&  0.22& 	0.03$-$0.05   \\
J1544$-$5308&0.03$-$0.05&0.02&  0.23& 	0.02$-$0.04   \\
J1559$-$4438&0.02$-$0.03&0.01&  0.15& 	0.01$-$0.02   \\
J1603$-$7202&0.03$-$0.05&0.02&  0.15& 	0.02$-$0.03   \\
J1605$-$5257&0.03$-$0.05&0.02&  0.24& 	0.03$-$0.04   \\
J1607$-$0032&0.05$-$0.12&0.06&  0.37& 	0.06$-$0.09   \\
J1614$+$0737&0.02$-$0.05&0.03&  0.12& 	0.03$-$0.04   \\
J1709$-$1640&0.02$-$0.05&0.03&  0.19& 	0.02$-$0.04   \\
J1713$+$0747&0.02$-$0.06&0.06&  0.12& 	0.04$-$0.06   \\
J1730$-$2304&0.05$-$0.12&0.07&  0.47& 	0.06$-$0.09   \\
J1744$-$1134&0.07$-$0.16&0.12&  0.55& 	0.09$-$0.14   \\
J1751$-$4657&0.03$-$0.06&0.03&  0.24& 	0.03$-$0.04   \\
J1752$-$2806&0.02$-$0.04&0.02&  0.16& 	0.02$-$0.03   \\
J1848$-$1952&0.03$-$0.06&0.05&  0.16& 	0.04$-$0.06   \\
J2053$-$7200&0.04$-$0.08&0.05&  0.14& 	0.05$-$0.06   \\
\hline
\end{tabular}
\end{center}
Note: Columns (2)--(5) list the fractional distances (e.g., \db=\Db/D). \\
~~~~~~$^{\ast}$ For the \Db range of the solid and dashed envelopes in Fig.~\ref{plotfour}. \\
~~~~~~$^{\star}$ For the geometry in Fig.~\ref{plotfour} (i.e., center located at $\sim$170 pc from the Sun, 
towards ($l,b$)=($329^{\circ},17.5^{\circ}$)) \\
\label{tabletwo}
\end{table}


\begin{table}
\caption{Various models considered and their goodness parameters}
\begin{center}
\begin{tabular}{lcccc}
\hline
\hline
       		& 	     &	       &	    &		\\
Model   	& Parameters & \epsall & \epslow    & \epshigh  \\
       		& 	     &	       &	    &		\\
\hline
\hline
LB		& \SMLB=0.2 + ambient ISM                  & 1.06  & 0.67  & 1.41 	\\
LB+LI		& \SMLB=0.2, \SMLI=0.29                & 0.55  & 0.06  & 0.90	\\
LB+LI+SA	& \SMLB=0.2, \SMLI=0.29, \cnsa=0.047   & 0.22  & 0.06  & 0.34	\\
LB-LI IZ-A	& \SMIZ=1.12, \Dbz=0.5(\Dbsmall+\Dbone)& 0.68  & 0.13  & 1.09	\\
LB-LI IZ-B	& \SMIZ=0.83, \Dbz=0.5(\Dblarge+\Dbone)& 0.65  & 0.11  & 1.06	\\
IZ-A+SA		& \SMIZ=1.12, \cnsa=0.047 	           & 0.27  & 0.13  & 0.38   \\
IZ-B+SA		& \SMIZ=0.83, \cnsa=0.047	           & 0.25  & 0.11  & 0.37   \\
\hline
\end{tabular}
\end{center}
\label{tablethree}
\end{table}

\end{document}